# Ferroelectric Capped Magnetization in Multiferroic PZT/LSMO Tunnel Junctions

**Ashok Kumar[1*], D. Barrionuevo[2], N Ortega[3], A. K. Shukla[1], Santiranjan Shannigrahi[3], J. F. Scott[4] and Ram S. Katiyar[2]**

[1]National Physical Laboratory (CSIR), Dr K S Krishnan Road, New Delhi-110012, India

[2]Department of Physics and Institute of Functional Nanomaterials, University of Puerto Rico, San Juan, Puerto Rico 00931-3343, USA

[3]Institute of Materials Research and Engineering (IMRE), 3 Research Link, Singapore 117602

[4]Department of Chemistry and Department of Physics, University St. Andrews, St. Andrews KY16 ST, UK

**Abstract:**

Self-poled ultra-thin ferroelectric $PbZr_{0.52}Ti_{0.48}O_3$ (PZT) (5 and 7 nm) films have been grown by pulsed laser deposition technique on ferromagnetic $La_{0.67}Sr_{0.33}MnO_3$ (LSMO) (30 nm) to check the effect of polar capping on magnetization for ferroelectric tunnel junction (FTJ) devices. PZT/LSMO heterostructures with thick polar PZT (7 nm) capping show nearly 100% enhancement in magnetization compared with thin polar PZT (5 nm) films, probably due to excess hole transfer from the ferroelectric to the ferromagnetic layers. Core-level X-ray photoelectron spectroscopy studies revealed the presence of larger Mn 3s exchange splitting and higher $Mn^{3+}/Mn^{4+}$ ion ratio in the LMSO with 7 nm polar capping.

*Authors to whom correspondence should be addressed. Electronic addresses: ashok553@nplindia.org

Highly polar ferroelectric thin (< 8 nm) films capable of tunneling charge carriers sandwiched between metal-ferromagnetic-metal electrodes produce ferroelectric tunnel junctions (FTJ).[1,2,3] Advancements in thin-film growth technologies make it possible to fabricate epitaxial, defect-free, relaxed or strained polar thin films suitable for tunneling devices. FTJs have been a fascinating area of research due to their potential applications as multi-state nonvolatile memory elements with fast write and read logic bits, low power consumption, and small heat dissipation.[4] In addition to using normal metal electrodes in FTJs, researchers are also working with ferromagnetic metal electrodes to make ferroelectric (FE)-ferromagnetic (FM) multiferroic heterostructures that provide an extra degree of freedom to manipulate the logic states. These multiferroic heterostructures provide extra logic states but at the cost of more complex densities of states for charge carriers crossing the junctions compared to normal metal electrodes; hence a careful study across the FE/electrode interface is needed.[5,6]

One should understand the major problems across the interfaces of the FTJs. These are as follows: (i) less than 3-nm thickness of ferroelectric capping shows poor ferroelectricity for direct transport measurements, using photolithography or electron beam lithography and device areas of a few microns; (ii) quantum tunneling effects diminish greatly for films thicker than 8 nm. Hence, to maintain both good ferroelectricity and strong tunneling effects in heterostructures, a thickness (d) window exists only for 3 nm < d < 8 nm. The present study therefore deals with polar PZT (5 nm and 7 nm) capping on 30 nm thick LSMO films. [7,8]

Several experimental and theoretical research works on FTJ have been reported based on ferroelectric polar barriers such as Pb(Zr,Ti)O$_3$ [9,10, 11], BaTiO$_3$ [12,13], (Ba,Sr)TiO$_3$ [14] and polyvinylidene fluoride (PVDF) (70%)–trifluoroethylene (TrFE) (30%) [15]. In these devices polarization reversal provides distinct and significantly different resistive states due to tunnel magnetoresistance (TMR) and tunnel electroresistance (TER). PZT has a track record of good ferroelectricity down to 2.4 nm and it is popular among researchers for studies based on ultra-thin films.[16] Among strongly correlated materials, strontium-doped lanthanum manganite (LSMO) has shown colossal magneto-resistance phenomena under application of small external magnetic fields. Exchange among the various valence states of Mn-cations leads to complex magnetization properties and transport behavior of LSMO, modulated by external hole/electron doping and electric/magnetic field application. PZT, LSMO, and their heterostructures are of special interest due to rich basic physics and technological applications. There are several reports on the x-ray photoelectron spectroscopy (XPS) analysis of PZT and LSMO thin films which deal with the compositional analysis of both systems and the valence states of each individual element. A few XPS reports are also available in the literature on PZT/LSMO heterostructures; however a thick layer of PZT restricts the investigation of chemical and valence states of layers close to interface. [17,18]

Preziosi et al. [10] have carefully demonstrated the transport properties of LSMO (5 nm) under two different polar states of PZT, which revealed that ferroelectric polarization significantly changes the interfacial magnetic anisotropy and affects the spin-orbit coupling. Leufke et al. [11] have shown significant modification of magnetization of LSMO (7.2 nm) thin films with thick polar PZT (90 nm) layers. Switching of ferroelectric polarization towards the LSMO surface (downward) significantly increased the magnetization below the Curie temperature ($T_c\sim$ 247 K for 8-nm LSMO film). In this situation holes accumulate at the LSMO interface, which significantly increases the magnetization; and opposite n-type state occurs for upward direction of polarization. As expected, this situation reverses above $T_c$, which qualitatively differs only in the sign of the modulation.

The presence of ferroelectric and magnetic dead layers is well known in the literature; their thicknesses were found to be 2-5 nm in manganite compounds and 1-2 nm in ferroelectric perovskites. The dead-layer effect is also crucial in the magnetic and polarization properties of FTJs and in magnetic tunnel junction (MTJ) devices which deal with ultra-thin films.[19] In most investigations the active layers of heterostructures are much thicker than the dead layers; only a few reports on multiferroic heterostructure field-effect devices even mention the existence of dead layers. Brivio et al. [20]report the absence of any electric field effect on magnetic modulation of 3-nm LSMO thin film when it was bottom-gated with non-polar $SrTiO_3$ (STO) thin film, however the effect was significant when applied from the top side. They proposed that this effect was due the presence of a dead layer at the bottom STO/LSMO interface. Molegraaf et al. [21]illustrate a shift of nearly 20 K in the Curie temperature and a 20 % change in magnetic modulation for STO/LSMO/PZT heterostructures and a change in sign of modulation near 150 K.

Magnetic control of ferroelectric polarization and perturbation of the first unit cell of LSMO under applied electric field, probed by magnetic second-harmonic generation, were shown on PZT/LSMO heterostructures. [22,23] We have recently demonstrated the tunneling electroresistance (TER) with large resistance ratios (up to 100:1) in hetrostructures which mainly depends on the active device area. [24] The effect of magnetic fields on TER was also observed and was much greater with applied in-plane magnetic field. Here we report that the capping of polar ferroelectric PZT on LSMO significantly modulates the magnetization of LSMO. The magnetic properties of manganites strongly depend on the Mn valences and the Mn-O-Mn bond angle. XPS studies have been carried out to understand the underlying physics responsible for magnetic modulation.

The PZT/LSMO heterostructure films were grown on $(LaAlO_3)_{0.3}(Sr_2AlTaO_6)_{0.7}$ (LSAT) (001)-oriented substrates, using a pulsed laser deposition (PLD) system. The PZT and LSMO were grown at 600 °C and 700 °C respectively, under an oxygen pressure of ~80 mTorr, using a KrF excimer laser ($\lambda$ = 248 nm) with energy density of ~1–2 J cm$^{-2}$ and deposition frequency of 1 Hz, followed by annealing at 700 °C for 30 min in oxygen at a pressure of ~320 Torr. Surface morphology, growth process and thickness were monitored by an *in situ* reflection high-energy electron diffraction (RHEED) system operating at 21 keV beam energy, 1.4 A filament current and incident angles of 1°–4°. The crystal structure of the films was characterized using x-ray diffraction (XRD) technique. A multimode Nanoscope V (Veeco Instruments) was used to examine the morphology and piezoelectric properties using a conducting tip. The details of electrical characterization results are described elsewhere. [24] The XPS measurements were carried out with a commercial VG ESCALAB 220I-XL imaging system. The XPS spectra were obtained using an Al K$_\alpha$ x-ray source (1486.6 eV) and photoelectrons were analyzed by a hemispherical mirror energy analyzer. C 1s (285 eV) core level is used for binding energy calibration and charging correction. Shirley method has been used to subtract the background from the raw data. The magnetic properties were measured using the physical property measurement system (PPMS; Quantum Design). The magnetization versus temperature (M-T) measurements were carried out in field-cooled (FC) mode where the samples were first cooled to 10 K under 100 and 1000 Oe magnetic fields, then warmed to 300 K to obtain the FC data. After the M-T measurements were made, hysteresis curves were measured at 50 K under applied fields up to 1 T.

Systematic XRD studies have been performed on PZT/LSMO/LSAT heterostructures from 100 nm to down 3 nm PZT with constant LSMO thickness. [24,25] It was found that 2θ shifted to lower angles with decrease in film thickness. Fig. 1 shows the weak XRD patterns of both PZT 5 and 7 nm samples, oriented along (001/002) planes. The lattice parameters of bulk LSAT, LSMO and tetragonal PZT are $a_{LSAT}$=3.868 Å, $a_{LSMO}$=3.871 Å, and ($a_{PZT}$=4.036 Å, $c_{PZT}$=4.146 Å), respectively. These observations suggest a large in-plane compressive strain (-4.26 %) developed in PZT, especially in ultra-thin films, during the growth process, as can be seen in XRD patterns (2θ for PZT ~ 44.56° and LSMO/LSAT ~ 47.16°). The inset of Fig. 1 illustrates a three-dimensional (3D) image (3x3 μm$^2$) of the polarization switching for both PZT films using piezoresponse force microscopy (PFM) with +/- 3 V bias voltages. A distinct change in the contrast of poled PZT film surfaces can be seen with positive and negative E-field poling. The PFM scan was first performed over a 2x2 μm$^2$ area with +3 V tip bias, later at the center over 1x1 μm$^2$ with -3 V tip bias, and finally over the 3x3 μm$^2$ area without any tip voltage. The contrast of the -3 V tip bias poling area well matched with the contrast of the area without any poling. These observations imply that as-grown samples are self-poled towards the surface of LSMO with switchable polarization.

Fig. 2 (a, b) show the magnetic moments of PZT (5 and 7 nm) /LSMO (30 nm) /LSAT heterostructures recorded over 25 to 300 K under 100 Oe and 1000 Oe magnetic fields, below and above the coercive field. M-T responses show similar trends using 100 and 1000 Oe fields with greater saturation for 1000 Oe. The magnitude of the magnetic moment of 7-nm PZT-capped LSMO is almost 100% higher than that for 5-nm ferroelectric capping; note that the magnetic properties of these films come from similar LSMO dimensions (0.5 cm x 0.5 cm x 30 nm). The magnetic phase transition of LSMO for 5-nm ferroelectric capping was 275 (+/- 5 K), calculated using Arrott's method [26], far lower than that of the 7-nm PZT-capped LSMO film. The Curie temperature for the 7-nm PZT-capped LSMO was found to be beyond the experimental limits. Temperature-dependent differential magnetization (inset Fig2. (b)) exhibited kink for the 5-nm PZT-capped film; however, it was absent for 7-nm capped LSMO. Fig. 3 illustrates the comparative four quadrant M-H curve of both samples at 50 K, which indicates almost twice the remanent magnetization for the 7-nm capped PZT/LSMO compared to that of the 5-nm PZT; however, the coercive field in both cases was the same. The inset of Fig. 3 shows the large diamagnetic effect of the substrate for high probe field that is subtracted for M-H presentation. The first quadrant of the M-H curves indicates that the 5-nm PZT-capped LSMO was almost self- aligned; however, the 7-nm PZT-capped LSMO was less aligned and saturated for higher applied magnetic field.

To explain the magnetic anomaly, we have thoroughly investigated the chemical and compositional behavior of the PZT surfaces and the PZT/LSMO interfaces using XPS. XPS being a surface sensitive technique, it was difficult to probe the different elements of LSMO, since it is covered with PZT film. However, our choice of 5 and 7 nm thick PZT film still allowed us to acquire core-level XPS spectra, even though with weaker signal. Figure 4 (a) and (b) show the core-level XPS spectra of Pb $4p_{3/2}$ and Mn $2p_{3/2}$ of 5 and 7 nm PZT/LSMO respectively. It was difficult to analyze the spectra since Pb $4p_{3/2}$ peaks and Mn $2p_{3/2}$ peaks have overlapping BE region. The Pb 4p$_{3/2}$ peak related to PZT overlayers dominates the Mn 2p$_{3/2}$ peak originating from buried LSMO film. We have performed least $\chi^2$ fitting of experimental data using Gaussian components corresponding to Pb and Mn peak positions given in literature. [17, 18, 27] Deconvoluted components indicate that Mn$^{3+}$ ions are enhanced for 7 nm PZT-capped LSMO film compared to 5 nm PZT capped film. This enhancement in Mn$^{3+}$ ion related states may be due to the comparatively large hole doping by the self-poled 7 nm PZT film. To further support this proposal, we have analyzed the Mn 3s core-level spectra of 5 and 7 nm PZT capped LSMO film. In case of LSMO, Mn 3s core level is splitted in to two peaks due to exchange interaction of 3s photohole and unpaired 3d electrons. [28] Lower BE (majority spin character) and higher BE (minority spin character) exchange splitted peaks originate from parallel and anti-parallel coupling of remaining 3s electron spin and total spin of d-shell, respectively. Therefore, splitting of the Mn 3s can be taken as a measure of local magnetic moment of Mn atom. [29] We show Mn 3s core level spectra of 5 and 7 nm PZT capped LSMO film along with two deconvoluted fitting components and lower BE peak corresponding to exchange splitted peaks in Fig. 4 (c) and (d) respectively. We observe increase in the exchange splitting (energy separation between two peaks) for 7 nm PZT capped LSMO compared to 5 nm PZT capped LSMO which can be directly correlated to the enhanced Mn$^{3+}$/Mn$^{4+}$ ion ratio and consequent increase in the magnetic moment for 7 nm PZT capped LSMO film. [28]

Figure 5 (a) shows the Ti 2p XPS core-level spectra of 5-nm and 7-nm PZT-capped LSMO films, respectively. The two peaks observed are ascribed to Ti 2p$_{3/2}$ and Ti2p$_{1/2}$ spin-orbit components. [30] Raw data as well as fitted data do not reveal any shoulder on the higher BE side of Ti 2p$_{3/2}$ peak, which rules out any trace of Ti$^{3+}$ ions responsible for magnetism. Close up region Ti 2p$_{3/2}$ fitted peaks (inset of Fig. 5 (a)) demonstrate an energy shift of nearly 0.25 eV towards lower BE for 7 nm PZT-capped LSMO, which indicates relaxation in the in-plane compressive strain with increase in polar capping thickness. XPS spectra for Pb 4f states (Fig. 5 (b)) show almost negligible change in shape and broadening with increased thickness of capping layer; however, a 0.1-eV shift in peak position towards the lower BE side was observed for 7-nm PZT capping. Fig. 5 (c) shows the O 1s XPS core level spectra of 5 and 7 nm PZT/LSMO heterostructures. It would be unwise to discriminate the contribution of O 1s from PZT and LSMO; in general both systems show indistinguishable XPS peaks. O 1s core level has been fitted using two Gaussian components corresponding to oxygen signal from lattice (bulk) and surface contaminants ($OH^-$ and/or C-O). The only distinction in O 1s XPS spectra is the small shift to the lower BE side and less contribution of surface oxygen for 7-nm PZT capping on LSMO films. The analysis of XPS spectra implies the presence of less in-plane compressive strain in the 7-nm PZT-capped LSMO films.

In summary, this investigation provides special information to the researchers working in the area of FTJ with limited options for ferroelectric thickness. Self-polarized polar ferroelectric PZT ultra-thin films may significantly modulate the magnetization of the underlying magnetic layer. Magnetic moment of LSMO layer was doubled with increase in thickness of polar capping; and the Curie temperature shifted nearer that for bulk LSMO. XPS analysis revealed that 7-nm PZT capping generated excess holes from the polar layer to the magnetic LSMO layer produce larger Mn 3s exchange splitting and higher magnetic moment $Mn^{3+}$ / $Mn^{4+}$ ion ratio. In-plane compressive strain also decreased with increase in polar capping thickness; significantly modifying the PZT surface and PZT/LSMO interface.

**Acknowledgement:**

This work was supported by DOE (DE-FGD2-08ER46526) grant. D.B acknowledges support from IFN-NSF Grant NSF-RII-1002410 for the graduate fellowship.

**Figure captions**

Fig. 1. The XRD patterns of PZT (5 nm and 7 nm)/LSMO/LSAT heterostructures oriented along the (001/002) plane. Inset illustrates their 3D PFM images under (+/- 3V).

Fig. 2. M-T graphs of PZT /LSMO (30 nm)/LSAT heterostructures; (a) 5-nm PZT capping; (b) 7-nm PZT capping [at 100 and 1000 Oe magnetic field]. The inset of Fig. 2 (b) shows dM/dT as function of temperature at 1000 Oe field.

Fig. 3. M-H graphs of PZT (5 nm and 7 nm) /LSMO (30 nm)/LSAT heterostructures at 50 K; inset shows applied field dependence up to 1.0 T.

Fig. 4. (a-d) show the core-level XPS spectra of Pb $4p_{3/2}$, Mn $2p_{3/2}$, and Mn 3s peaks: (a) Pb $4p_{3/2}$, Mn $2p_{3/2}$ for 5-nm PZT capping; (b) Pb $4p_{3/2}$, Mn $2p_{3/2}$ for 7-nm PZT capping; (c) Mn 3s for 5-nm PZT capping; and (d) Mn 3s for 7-nm PZT capping

Fig. 5. (a-c) show the core level XPS spectra of: (a) Ti 2p, for 5-nm and 7-nm PZT capping (inset shows fitted spectra for comparison); (b) comparative Pb 4f peaks, and (c) O1s peaks for both 5- and 7-nm PZT capping. All the spectra have been normalized to same peak height for comparison and staggered vertically for clarity of presentation.

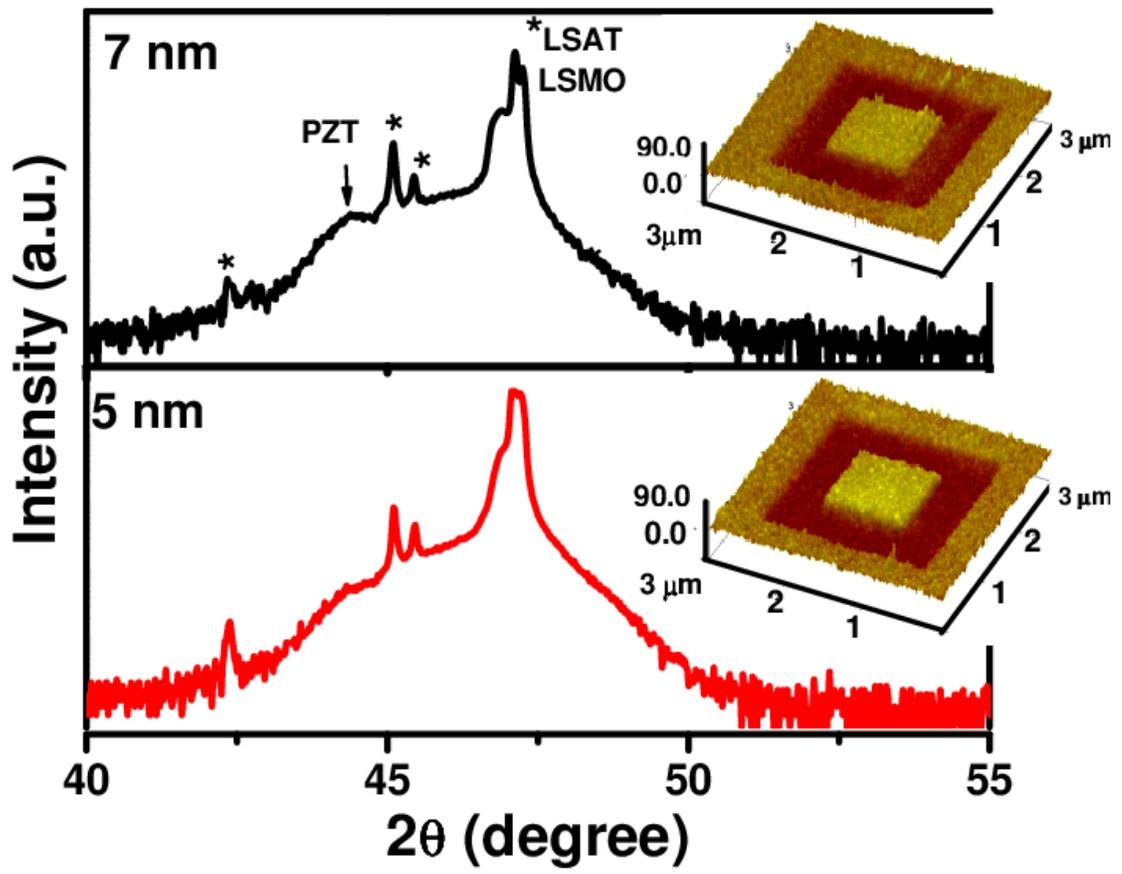

Fig.1

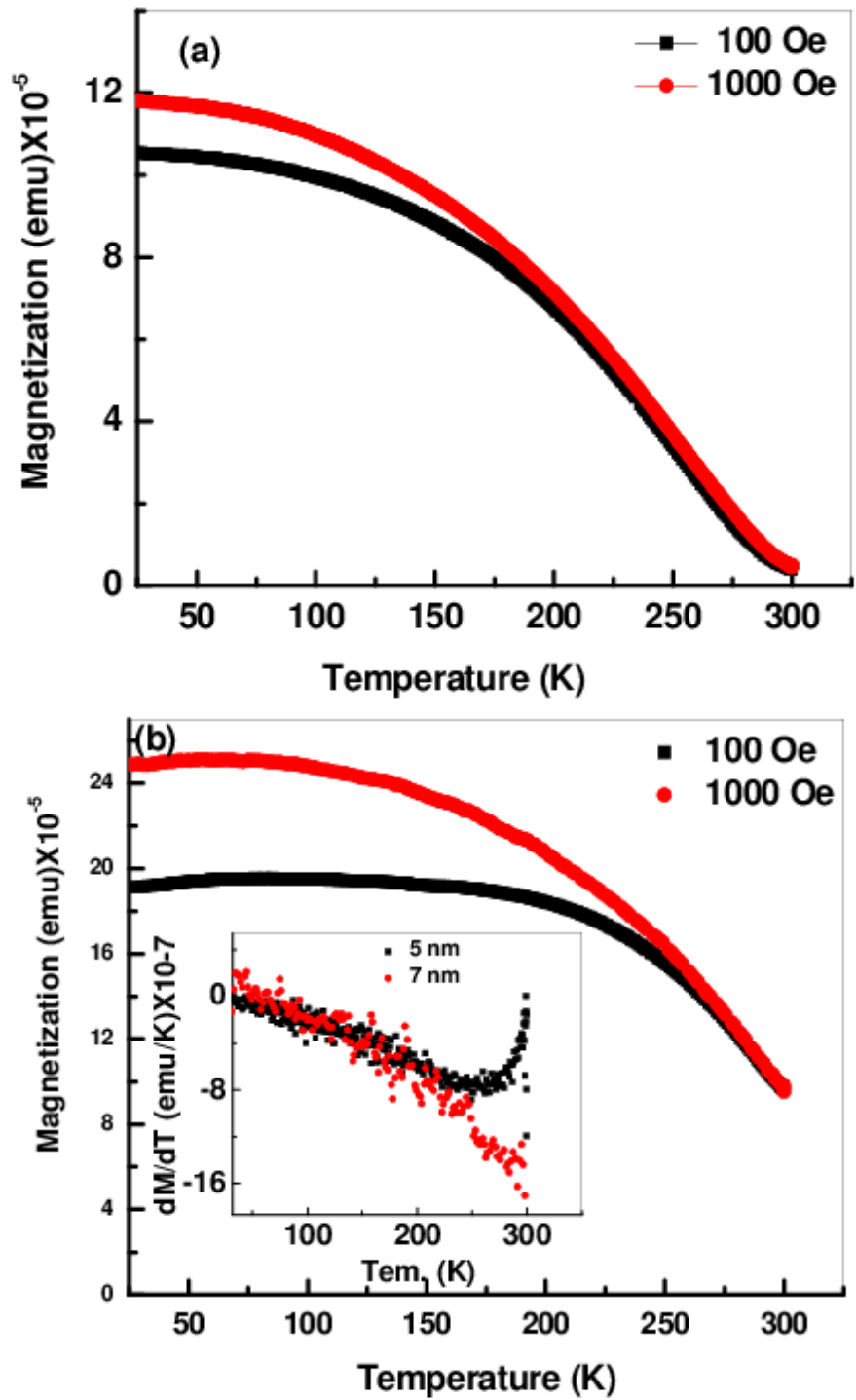

Fig.2

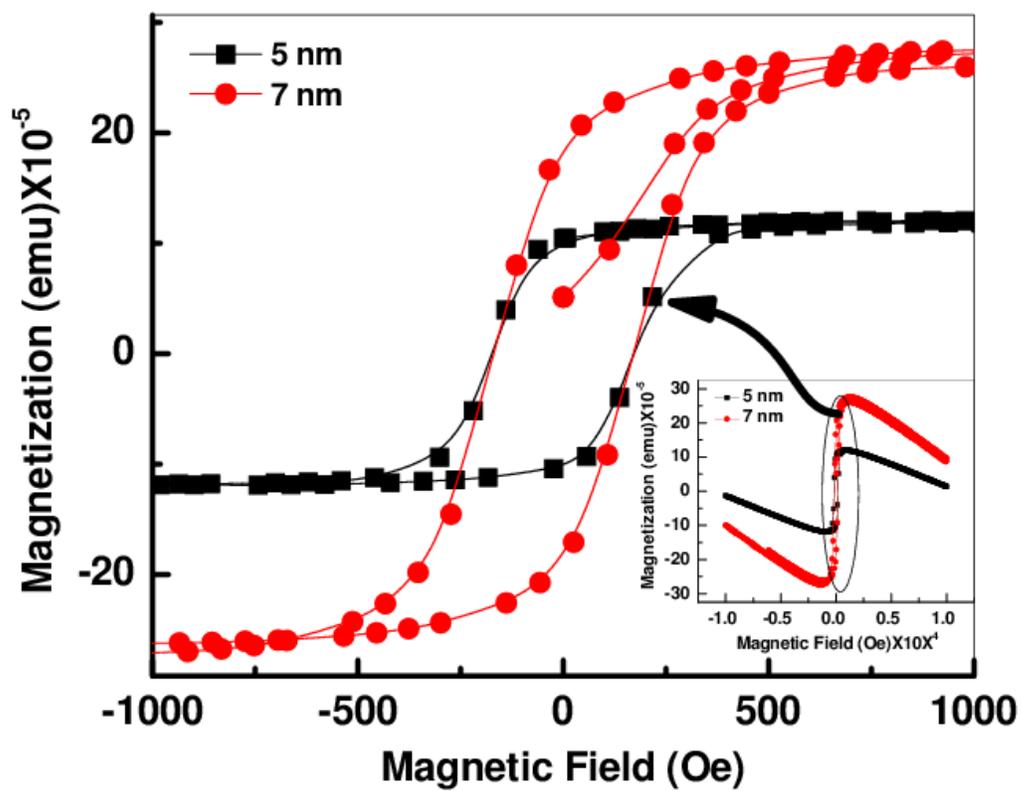

Fig.3

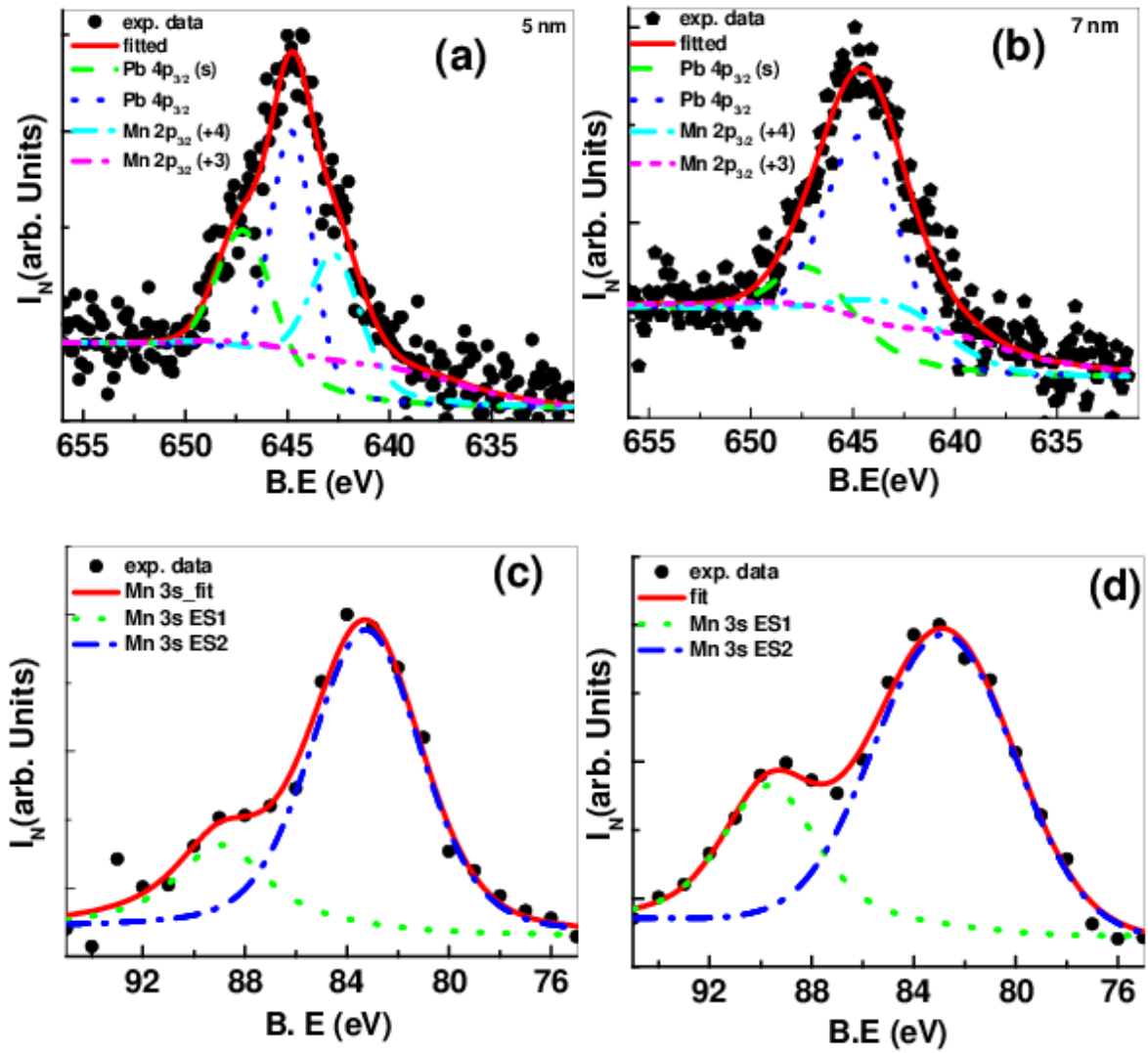

Fig.4

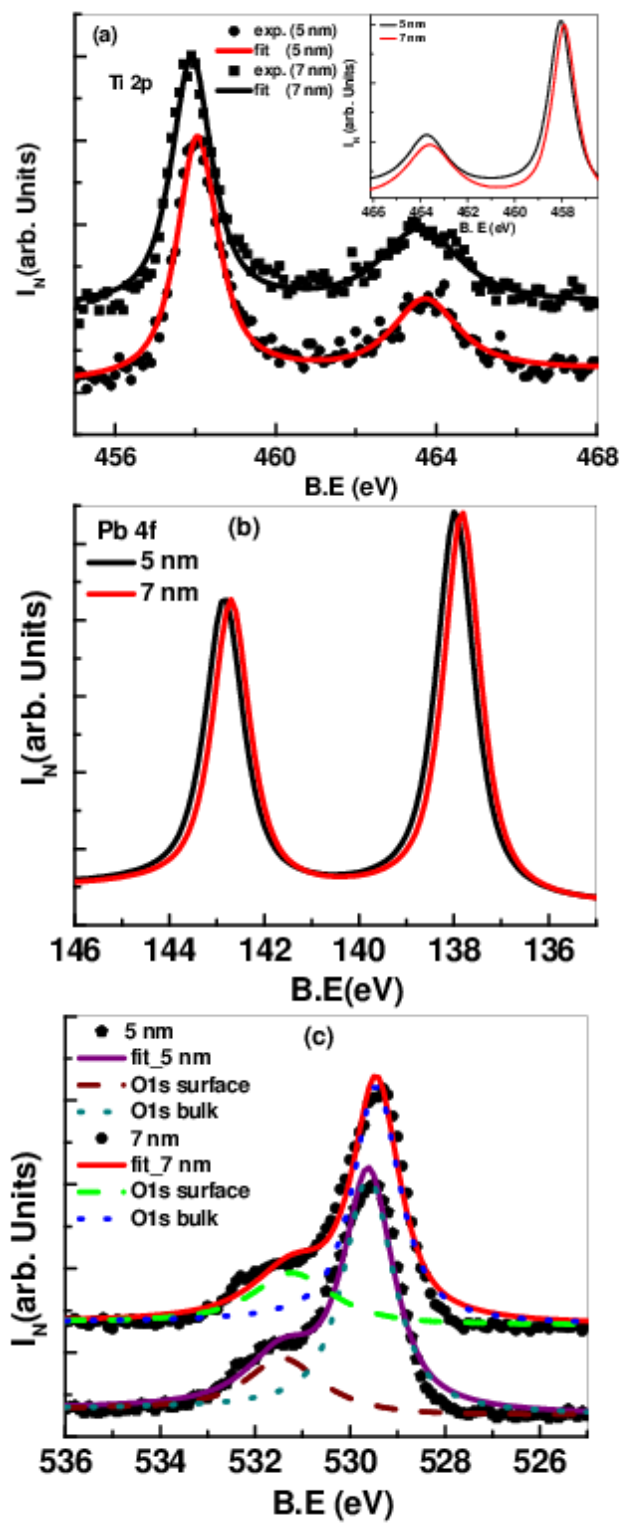

Fig.5